\documentclass[aps,prb,twocolumn]{revtex4-1}

\usepackage{graphicx}
\usepackage{color}
\usepackage{soul}
\usepackage{multirow}
\usepackage{amsmath}

\def\CHR{Cr$_2$O$_3$}

\begin{document}

\title{First-principles microscopic model of exchange-driven magnetoelectric response with application to Cr$_2$O$_3$}

\author{Sai Mu}
\affiliation{Department of Physics and
Astronomy and Nebraska Center for Materials and Nanoscience, University of
Nebraska-Lincoln, Lincoln, Nebraska 68588, USA}
\author{A. L. Wysocki}
\altaffiliation[Present address: ]{Ames Laboratory, Ames, Iowa 50011, USA.}
\affiliation{Department of Physics and Astronomy and Nebraska Center for Materials and Nanoscience, University of
Nebraska-Lincoln, Lincoln, Nebraska 68588, USA}
\author{K. D. Belashchenko}
\affiliation{Department of Physics and
Astronomy and Nebraska Center for Materials and Nanoscience, University of
Nebraska-Lincoln, Lincoln, Nebraska 68588, USA}

\date{\today}

\begin{abstract}
The exchange-driven contribution to the magnetoelectric susceptibility $\hat\alpha$ is formulated using a microscopic model Hamiltonian coupling the spin degrees of freedom to lattice displacements and electric field, which may be constructed from first-principles data. Electronic and ionic contributions are sorted out, and the latter is resolved into a sum of contributions from different normal modes. If intrasublattice spin correlations can be neglected, the longitudinal component $\alpha_\parallel$ becomes proportional to the product of magnetic susceptibility and sublattice magnetization in accordance with Rado's phenomenological model. As an illustration, the method is applied to analyze the temperature dependence of the longitudinal magnetoelectric susceptibility of Cr$_2$O$_3$ using first-principles calculations and the quantum pair cluster approximation for magnetic thermodynamics. A substantial electronic contribution is found, which is opposite to the ionic part. The sensitivity of the results to the Hubbard $U$ parameter and the sources of error are studied. It is also found that non-Heisenberg interactions are too weak to account for the sign change of $\alpha_\parallel$ in Cr$_2$O$_3$.
\end{abstract}

\maketitle

\section{Introduction}\label{intro}

Antiferromagnets exhibiting a linear magnetoelectric effect \cite{Landau} are of great interest for applications aiming to achieve electric control of magnetism. \cite{Fiebig,Borisov,Binek,Bibes,He} In such materials there is a term $- \mathbf{E}\hat\alpha\mathbf{H}$ in the free energy density, where $\hat\alpha$ is the magnetoelectric tensor. Due to this term the electric field induces a magnetization and the magnetic field induces a dielectric polarization, both in linear order. Magnetoelectric effect was first predicted \cite{Dzyaloshinskii} and experimentally observed \cite{Astrov,Folen,Astrov2,Rado} in \CHR, which remains the most promising material for applications.

Magnetoelectric effect can arise due to several microscopic mechanisms, including electric field-induced changes of the single-ion anisotropy, Heisenberg exchange parameters, the $g$-tensor or Dzyaloshinskii-Moriya interaction (see Ref.\ \onlinecite{Bonfim} for a review). Each of these contributions can be further divided into electronic (clamped-ions) and lattice-mediated parts. Experimentally the electronic contributions can, in principle, be measured separately at frequencies that are large compared to those of the relevant optical phonon vibrations.

First-principles methods can illuminate the microscopic mechanisms of the magnetoelectric effect.  \cite{Iniguez,Delaney,Wojdel,Wojdel2,Mostovoy,Prosandeev,Bosquet,Bosquet2,Malashevich,Scaramucci} At zero temperature it is controlled by spin-orbit coupling. \'I\~niguez \cite{Iniguez} showed that at zero temperature the lattice-mediated contribution can be obtained by evaluating the electric and magnetic polarities and stiffnesses of the polar displacement modes. An alternative approach \cite{Bosquet} is to compute the electric polarization induced by the magnetic field. This method was used to calculate both lattice-mediated and electronic contributions to the transverse magnetoelectric susceptibility $\alpha_\perp$ of \CHR. The electronic contribution turned out to be as much as one third of and have the same sign as the lattice-mediated one. \cite{Bosquet} The orbital contribution to the magnetoelectric response has also been considered. \cite{Malashevich,Scaramucci}

Longitudinal magnetoelectric susceptibility $\alpha_\parallel$ reaches a maximum at finite temperature, where it is dominated by Heisenberg exchange. This temperature-dependent effect in \CHR\ was studied by Mostovoy \emph{et al.} \cite{Mostovoy}, who obtained the relevant coupling constant from the electric polarization of a ferrimagnetically ordered unit cell and the temperature dependence from Monte Carlo simulations for the classical Heisenberg model. Their approach is tailored to \CHR\ and is not directly applicable to other systems. Only the total response was evaluated.

In this paper we formulate a microscopic model of exchange-driven magnetoelectric response which generalizes the approach of Ref.\ \onlinecite{Mostovoy}. We study the longitudinal magnetoelectric susceptibility of \CHR\ in more detail, sorting out the electronic and lattice-mediated contributions and resolving the latter by normal displacement modes, and comparing the predictions of the quantum pair cluster, quantum mean-field, and classical mean-field approximations. We also test the possibility that non-Heisenberg spin interactions could be responsible for the sign change of $\alpha_\parallel$ in \CHR\ and conclude in the negative.

The paper is organized as follows. In Section \ref{model} the microscopic model is formulated in terms of the microscopic coefficients coupling the spins to lattice displacements and directly to the electric field, and the general expressions for the electronic and lattice-mediated contributions to magnetoelectric susceptibility are derived. The computational procedure is described in Section \ref{method}, and the results are presented in Section \ref{results}. We find that the electronic contribution is a sizeble fraction of the lattice-mediated term and its sign is \emph{opposite}.
Different statistical approximations lead to similar maximal values of the magnetoelectric susceptibility, but the latter is sensitive to the choice of the Hubbard $U$ parameter due to its effect on the magnetic susceptibility. Section \ref{conclusions} concludes the paper.

\section{Microscopic model}\label{model}

Here we restrict ourselves to the exchange-driven magnetoelectric effect. This means that the model Hamiltonian in zero magnetic field should be invariant under a coherent rotation of all spins. This restriction is appropriate for Cr$_2$O$_3$ where the orbital moments are almost completely quenched, but it may need to be relaxed for application to other systems with strong spin-orbit coupling. A spin orientation for a magnetic atom on lattice site $i$ will be denoted by a unit vector $\mathbf{e}_i$. Spin rotation symmetry implies that $\mathbf{e}_i$ should only appear in scalar combinations. We start from the expansion of the effective Hamiltonian to second order in lattice displacements $\mathbf{u}_i$ from an equilibrium reference state:
\begin{align}\label{Ham}
    H&=E_0+\frac12\sum_{ij}\mathbf{u}_i\hat A_{ij}\mathbf{u}_j-\sum_i\mathbf{u}_i\hat
q^*_{i}\mathbf{E}-\mathbf{H}\sum_i\mu_i\mathbf{e}_i    \nonumber\\
    &+H_{mag}\{\mathbf{e}_i\}
-\frac12\sum_{jk}\left(\sum_i \mathbf{u}_i\mathbf{g}_{i,jk} + \mathbf{E}\mathbf{f}_{jk}\right)
(\mathbf{e}_j\mathbf{e}_k)
\end{align}
where $\mu_i$ is the magnetic moment at site $i$, $\mathbf{E}$ and $\mathbf{H}$ the external electric and magnetic fields,
$\hat q^*_i$ the Born effective charge tensors, $\hat A_{ij}$ the Born-von K\'arm\'an force constant matrix, and $H_{mag}$ the magnetic interaction Hamiltonian in the absence of displacements and external fields. The Hamiltonian (\ref{Ham}) is generally similar to that of Ref.\ \onlinecite{Yatom}, but we explicitly sort out the coupling of spins to lattice displacements in order to separate the ionic and electronic contributions to the magnetoelectric susceptibility.

The magnetoelectric coupling is generated by the last term in (\ref{Ham}), where the sum over $j$, $k$ runs over magnetic sites only. The parameters may be expressed as
\begin{equation}
\mathbf{g}_{i,jk} = \frac{\partial J_{jk}(\mathbf{u},\mathbf{E})}{\partial\mathbf{u}_i} \; , \quad
\mathbf{f}_{jk} = \frac{\partial J_{jk}(\mathbf{u},\mathbf{E})}{\partial\mathbf{E}}\label{coefs}
\end{equation}
where $J_{jk}$ is the Heisenberg exchange parameter for a pair of spins on sites $j$, $k$.
The vectors $\mathbf{f}_{jk}$ describe the variation of the exchange parameters when external electric field is applied while the ions are clamped; this term generates the electronic magnetoelectric response. The vector $\mathbf{g}_{i,jk}$ gives the variation of the exchange parameter for pair $jk$ when the site $i$ is shifted from the reference configuration, or it can conversely be interpreted as the spin-dependent Kanzaki force.\cite{Kanzaki} Translational invariance demands
\begin{equation}\label{sum-rule}
    \sum_i\mathbf{g}_{i,jk}=0.
\end{equation}

Coupling of atomic displacements to non-Heisenberg (e.\ g.\ biquadratic) spin interaction terms is also possible, but these terms are usually small in wide-gap insulators (this is explicitly checked below for \CHR). For systems where spin-orbit interaction has a significant contribution to the magnetoelectric response, the model may be extended by treating the exchange parameters as second-rank tensors. In this case the vectors $\mathbf{g}_{i,jk}$ and $\mathbf{f}_{jk}$ turn into third-rank tensors.

Integrating out the spin degrees of freedom and treating the external magnetic field and the last (magnetoelectric) term in
(\ref{Ham}) as small perturbations, we obtain an effective Hamiltonian for the lattice degrees of freedom:
\begin{align}\label{Heff}
    H_\mathrm{eff}=-\sum_i\mathbf{u}_i\hat q^*_{i}\mathbf{E}
    +\frac12\sum_{ij}\mathbf{u}_i\hat A_{ij}\mathbf{u}_j
    -\frac12\mathbf
    H\hat\chi^0_m\mathbf{H}\nonumber\\
    -\frac\beta2\sum_{jk}\left(\sum_i\mathbf{u}_i\mathbf{g}_{i,jk}+\mathbf{Ef}_{jk}\right)\sum_l\mu_l\left<(\mathbf{e}_j\mathbf{e}_k)\mathbf{e}_l\right>_0\mathbf{H}
\end{align}
where $\beta=1/kT$, $\hat\chi^0_m$ is the magnetic susceptibility tensor, $\left<\dots\right>_0$ the statistical average taken over the unperturbed state described by $H_{mag}$, and we have dropped the first-order magnetostrictive term generated by $\mathbf{g}_{i,jk}$, which does not affect the magnetoelectric response.

Let us denote $\mathbf{L}_{jk}=\frac12\sum_l\mu_l\left<(\mathbf{e}_j\mathbf{e}_k)\mathbf{e}_l\right>_0$ and
\begin{equation}\label{gi}
\hat g^*_i=\beta\sum_{jk}\mathbf{g}_{i,jk}\otimes\mathbf{L}_{jk}.
\end{equation}
Referring to (\ref{Heff}), we see that $\hat g^*_{i}\mathbf{H}$ represents the force acting on site $i$ arising in response to the external magnetic field $\mathbf{H}$. Thus, $\hat g^*_i$ is the \emph{effective magnetic monopole charge} of the atom at site $i$. Conversely, $\hat g^*_i$ describes the change in the magnetization arising due to the displacement of site $i$. It depends on temperature and is non-zero for both magnetic and non-magnetic atoms. The sum $\sum_i\hat g^*_i=0$ thanks to (\ref{sum-rule}).
The structure of the tensor $\hat g^*_i$ is determined by the corresponding magnetic site symmetry. For example, the symmetry of  the Cr sites in Cr$_2$O$_3$ includes the $C_3$ axis, and therefore for these sites only the $zz$ component of $\hat g^*_i$ is non-zero.

When a material is both piezoelectric and piezomagnetic, there is a contribution to the magnetoelectric susceptibility mediated by strain. Although this contribution can be derived from the same Hamiltonian (\ref{Heff}), we assume in the following that it is not present, as in the case of \CHR. Thereby we can treat $\mathbf{u}_i$ as internal to the unit cell and disregard the possible change of the volume and unit cell shape. Equilibrium displacements are found by minimizing (\ref{Heff}), which leads to
\begin{equation}\label{equil}
    \sum_j\hat A_{ij}\mathbf{u}_j=\hat q^*_i\mathbf{E}+\hat g^*_i\mathbf{H}.
\end{equation}

Since the displacements $\mathbf{u}_i$ are identical in all unit cells, in Eq.\ (\ref{equil}) we can treat $i$ and $j$ as basis indices within the unit cell and $\hat A_{ij}$ as the Fourier transform of the force constant matrix at $\mathbf{q}=0$.
This matrix has three zero eigenvalues corresponding to homogeneous lattice displacements and $3(N-1)$ finite
eigenvalues (where $N$ is the number of sites in the unit cell). Homogeneous displacements should be excluded from consideration, and we therefore consider only the subspace of displacements for which $\sum_i\mathbf{u}_i =0$. Within this subspace the action of the symmetric matrix $\hat A$ is represented by its eigendecomposition $A^{\mu\nu}_{ij}\to\sum_n V^n_{i\mu} C_n V^n_{j\nu}$, where $C_n$ are the non-zero eigenvalues and $V^n_{i\mu}$ the normalized eigenvectors. Within this subspace (with
homogeneous displacements projected out) the action of the matrix $\hat A$ can be inverted, resulting in $\mathbf{u}_i=\sum_j\hat G_{ij}\mathbf{F}_j$, where $\mathbf{F}_j$ are the forces in the right-hand side of (\ref{equil})
and $G^{\mu\nu}_{ij}=\sum_n V^n_{i\mu} C_n^{-1} V^n_{j\nu}$. The forces automatically satisfy the sum rule $\sum_i\mathbf{F}_i=0$, therefore the matrix $\hat G$ acts within its domain of definition.

Substituting the equilibrium displacements in (\ref{Heff}), we find the free
energy density:
\begin{align}\label{res}
    \frac{F}{V}=-\frac12\mathbf{E}\hat\chi_e\mathbf{E}-\frac12\mathbf{H}(\hat\chi^0_m+\delta\hat\chi_m)\mathbf{H}
    -\mathbf{E}\hat\alpha\mathbf{H}
\end{align}
where
\begin{align}\label{suscepte}
    \hat\chi_e&=\frac1\Omega \hat q^*_i\hat G_{ij}\hat q^*_j\\
    \delta\hat\chi_m&=\frac1\Omega \hat g^*_i\hat
    G_{ij}\hat g^*_j\label{susceptm}\\
    \hat\alpha&=\hat\alpha_{ion} + \hat\alpha_{el}\label{susceptme}\\
    \hat\alpha_{ion}&=\frac1\Omega \hat q^*_i\hat G_{ij}\hat g^*_j\label{a-ion}\\
    \hat\alpha_{el}&=\frac{\beta}{\Omega}\sum_{jk}\mathbf{f}_{jk}\otimes\mathbf{L}_{jk}\label{a-el}
\end{align}
Here $\Omega$ is the volume of the unit cell, $\hat\chi_e$ is the standard expression for the lattice-mediated dielectric susceptibility in the Born-von K\'arm\'an model, $\delta\hat\chi_m$ is the magnetostructural correction to the magnetic susceptibility, and $\hat\alpha$ is the magnetoelectric tensor, which includes the ionic (first) and the electronic (second) terms. We have dropped the phonon part of the free energy, which does not contribute to the linear susceptibilities in the harmonic approximation.

Denoting the dielectric and magnetic monopole polarities of an eigenmode $n$ as $\mathbf{p}_n=\sum_i\hat q^*_i\mathbf{V}^n_i$ and $\mathbf{g}_n=\sum_i\mathbf{V}^n_i\hat g^*_i$, we can rewrite the ionic part of the magnetoelectric tensor as
\begin{equation}\label{me}
    \hat\alpha_{ion}=\frac1\Omega\sum_nC_n^{-1}\mathbf{p}_n\otimes\mathbf{g}_n
\end{equation}
This expression agrees with that obtained by I\~niquez \cite{Iniguez} at zero temperature (where the monopole charges are controlled by spin-orbit coupling), but the present approach also provides a microscopic definition (\ref{gi}) of the temperature-dependent monopole charge.

The microscopic model based on the Hamiltonian (\ref{Ham}) can be generalized to the case of magnetoelectric alloys by adding configuration-dependent Kanzaki forces (both spin-independent and spin-dependent) and force constants. The parameters can be fitted to first-principles calculations of the total energies or forces using several supercells with different configurational and magnetic orderings and lattice displacements. Fitting of the parameters $\mathbf{f}_{jk}$ requires the calculation of the electric polarization at zero lattice displacements. The magnetoelectric response of a random alloy can then be obtained as an explicit configurational average of Eqs.\ (\ref{a-ion})-(\ref{a-el}).

The expressions (\ref{susceptme})-(\ref{me}) are valid for any magnetic structures, including multisublattice and noncollinear ones. They can be simplified in the case of a collinear antiferromagnet by noting that $\mathbf{L}_{jk}$ is parallel to the ordering axis and vanishes if sites $j$ and $k$ belong to equivalent antiferromagnetic sublattices with opposite magnetizations. Indeed, in this case the magnetic space group must contain a symmetry operation that interchanges the sites $j$ and $k$. This operation maps an arbitrary site $l$ onto an equivalent site with an opposite magnetization. The sum over $l$ is therefore zero. In the case of a two-sublattice antiferromagnet such as Cr$_2$O$_3$, it means $\mathbf{L}_{jk}\neq0$ only if $j$ and $k$ belong to the same sublattice. Taking these properties into account, we can define a scalar $l_{jk}$ as $\mathbf{L}_{jk}=\frac12\left<\mathbf{e}_j+\mathbf{e}_k\right>_0l_{jk}$, and $l_{jk}$ has the symmetry of the non-magnetic space group.

The magnetoelectric response generally depends in a complicated way on all the parameters appearing in the Hamiltonian (\ref{Ham}). A significant simplification can be achieved if it is admissible to neglect intrasublattice spin correlations. This approximation is justified as long as the corresponding exchange parameters $J_n$ satisfy $\beta J_n\ll1$. Intrasublattice interactions in antiferromagnets are usually not dominant, unless there is strong frustration. Therefore, the above inequality is satisfied in most antiferromagnets at temperatures that are not too low compared to $T_N$. In particular, this is the case for Cr$_2$O$_3$ where the shortest intrasublattice spin pair is the fourth-neighbor one.

The influence of weak interactions can be safely included on the mean-field level, while short-range interactions can still be treated using more accurate methods such as the thermodynamical cluster approximations. The neglect of correlations between sites $j$ and $k$ in the expression for $\mathbf{L}_{jk}$ leads, through a decoupling of spin averages, to $\beta l_{jk}\approx\chi$, where $\chi$ is the homogeneous longitudinal magnetic susceptibility,
\begin{equation}
\hat g^*_i \approx \chi m \sum_{jk} \mathbf{g}_{i,jk}\otimes \hat z \eta_{jk}
\end{equation}
where $m$ is the sublattice magnetization and $\eta_{jk}=+1$ ($\eta_{jk}=-1$) when $j$ and $k$ both belong to the sublattice with $m_z>0$ ($m_z<0$), and $\eta_{jk}=0$ if $j$ and $k$ belong to different sublattices, and finally
\begin{equation}\label{gn}
\mathbf{g}_n \approx \chi m \frac{\partial\Delta}{\partial u_n}\hat z,
\end{equation}
which should be substituted in Eq.\ (\ref{me}). In this last formula, $u_n$ is the amplitude of the $n$-th normal mode defined as $\mathbf{u}_i=u_n\mathbf{V}^n_i$, and $\Delta=(J^A_0-J^B_0)/2$, where $J^p_0=\sum_j J_{ij}$ with site $i$ belonging to sublattice $p$. Under the same condition the electronic contribution can be written as
\begin{equation}\label{alpha-el}
\alpha^{zz}_{el}=\frac1\Omega \chi m \frac{\partial\Delta}{\partial E_z}.
\end{equation}

Eqs.\ (\ref{gn})-(\ref{alpha-el}) have a structure equivalent to the phenomenological result of Ref.\ \onlinecite{Rado}. Thus, we see that this phenomenological form is appropriate for the exchange-driven magnetoelectric effect under the assumptions specified in the derivation of Eq.\ (\ref{me}). In particular, it does not require that the mean-field approximation is valid, but only that intrasublattice spin correlations are small, which is a weaker assumption.

\section{Computational procedure}\label{method}

\CHR\ is an antiferromagnetic insulator with a N\'eel temperature $T_N$ of 307 K. It has a corundum structure with the rhomboedral unit cell containing four equivalent Cr ions lying on the trigonal axis. The orientations of the Cr magnetic moments alternate along the trigonal axis. The magnetic point group $\bar 3'm'$ allows the magnetoelectric susceptibility tensor, which is diagonal and has two independent components, $\alpha_{\parallel}\equiv\alpha_{zz}$ and $\alpha_{\perp}\equiv\alpha_{xx}=\alpha_{yy}$, where the $z$ axis is aligned with the 3-fold axis. \cite{Borovik}  The $\alpha_\parallel$ component, which is dominated by exchange mechanism, is the focus of our study.

First-principles calculations were performed using the projector augmented wave method \cite{Blochl} implemented in the Vienna ab initio simulation package (VASP)\cite{Kresse,Kresse2}. The correlations within the Cr 3$d$ shell were described using the rotationally invariant LDA+U method. \cite{Liechtenstein} We set the Hund exchange parameter $J=0.58$ eV as obtained from the constrained occupation calculation, \cite{Shi} and studied the results as a function of the Hubbard $U$ parameter.

The energy cutoff for the plane wave expansion was set to 520 eV, and a $\Gamma$-centered Monkhorst-Pack $k$-point grid \cite{Monkhorst} was used for the Brillouin zone integration. Relaxations, phonon and Berry phase calculations were performed for the rhombohedral unit cell using 0.02 eV Gaussian smearing and a $8\times8\times8$ $k$-point mesh. The Hellmann-Feynman forces were converged to 0.005 eV/\AA. The exchange parameters were obtained using the hexagonal supercell and the tetrahedron method with the $4\times4\times2$ $k$-point mesh.

The $\mathbf{q}=0$ component of the force constant matrix is evaluated using the standard technique, and its non-uniform eigenvalues and eigenvectors are found. The Born effective charges are also calculated and used to evaluate the polarities $\mathbf{p}_n$ of the eigenmodes of the force constant matrix. Only the polar modes with nonzero $p_{nz}$ need to be considered.

The magnetic monopole charges of the normal modes in Eq.\ (\ref{gn}) contain two factors: $\chi m$, which is determined by $H_{mag}$, and $\partial_n\Delta=\partial\Delta/\partial u_n$. The unperturbed magnetic Hamiltonian $H_{mag}$ is assumed to have a Heisenberg form, and the exchange parameters are obtained by fitting the calculated total energies of a number of magnetic configurations. \cite{Shi}  The factors $\Delta_n$ are found by calculating $\Delta=(E_A-E_B)/4$, where $E_A$ and $E_B$ are the energies required to reverse one spin on sublattice A or B in the magnetic ground state. This is done using a 30-atom supercell with atomic displacements proportional to $\mathbf{V}^n_i$. Then $\Delta_n$ is found by numerical differentiation with respect to $u_n$.
This approach can be viewed as a converse of that used by Mostovoy \emph{et al.} The quantity $\partial\Delta/\partial E_z$ needed for the evaluation of the electronic contribution (\ref{alpha-el}) is calculated at zero atomic displacements in the presence of external electric field \cite{Souza} with a subsequent numerical differentiation.

\section{Results}\label{results}

There are two displacement modes with non-zero polarities $p_{nz}$, both transforming according to the $A_{2u}$ irreducible representation of the $\bar 3m$ point group. We denote these modes as LO$_1$ and LO$_2$. The normalized eigenvectors are listed in Appendix A, which also includes the phonon frequencies. The stiffnesses $C_n$ and dielectric polarities $p_{nz}$ of these modes are listed in Table \ref{number}, along with the values of the derivative $\partial_n\Delta$ which enters the expression (\ref{gn}) for the magnetic polarity. The stiffer LO$_1$ mode has a much larger dielectric polarity, which results in larger displacements compared to the softer LO$_2$ mode. The values of $\partial_n\Delta$ for the two modes are similar. Thus, overall the LO$_1$ mode gives a much larger contribution to $\alpha_\parallel$, which is seen from its larger value of $\partial\Delta/\partial E$. The electronic contribution to $\partial\Delta/\partial E$ is also listed in Table \ref{number}. It is comparable in magnitude to the lattice-mediated contribution, but the sign is opposite. We can understand this sign difference by noting that the electric field tends to perturb the electronic charge density in a way that partially compensates the displacement of the positively charged Cr ions. A significant magnetoelectric response was observed in optical measurements, \cite{Pisarev,Krichevtsov1,Krichevtsov2} but the sign of this response was not determined.

\begin{table}[htb]
\caption{Stiffnesses $C_n$, dielectric polarities $p_n$, and exchange perturbations $\Delta/E$ (see text) of the two polar displacement modes contributing to $\alpha_\parallel$ in \CHR, calculated at $U=4$ eV.}
\begin{tabular}{|c|c|c|c|c|}
\hline
Mode          & $C_n$ (eV/\AA$^2$)      & $p_{nz}$ ($e$) & $\partial_n\Delta$ (meV) & $\partial\Delta/\partial E$ (10$^{-3}e\cdot$\AA) \\
\hline
LO$_1$      &  29.1                                &   8.42                &    65.2        & 19.2      \\
LO$_2$      & 11.2                                 &   0.88                &     52.5         & 4.2       \\
Electronic   & ---                                    &   ---                   &   ---               & $-9.0$      \\
\hline
\end{tabular}
\label{number}
\end{table}

For the magnetic thermodynamics, which determines the factor $\chi m$ in (\ref{gn}) and (\ref{alpha-el}), we use the quantum pair cluster approximation \cite{Vaks} for $S=3/2$ and compare its predictions with quantum ($S=3/2$) and classical mean-field approximations. Since the corundum lattice is low-coordinated, we expect that the pair cluster approximation can provide a notable improvement compared to MFA due to the inclusion of short-range order effects, but at very low temperatures it breaks down by developing unphysical features. \cite{Vaks} The application of the pair cluster approximation is similar to Ref.\ \onlinecite{Shi}, with the exception that here we only treat the nearest and next-nearest neighbors within the pair-cluster approximation, while more distant pairs are included on the mean-field level. This is consistent with the neglect of the intrasublattice correlations in (\ref{gn}) and (\ref{alpha-el}) and is justified by the small magnitude of the exchange parameters beyond the second coordination sphere.

\begin{figure}[htb]
\includegraphics*[width=0.45\textwidth]{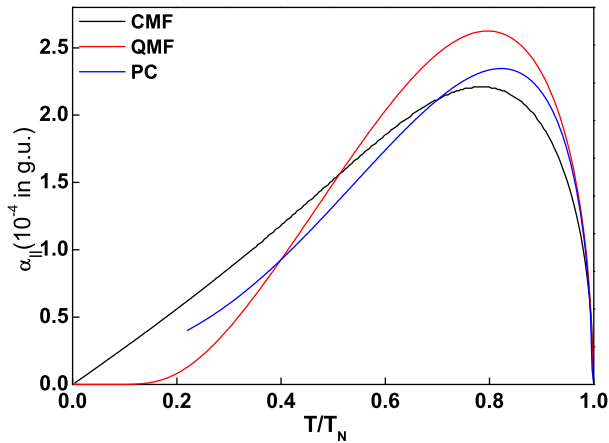}
\caption{Magnetoelectric susceptibility $\alpha_{\parallel}$ calculated using different statistical approximations: pair-cluster (PC), quantum mean-field (QMFA), and classical mean-field (CMFA).}
\label{methods}
\end{figure}

The temperature dependence of $\alpha_\parallel$ obtained using different statistical approximations is shown in Fig.\ \ref{methods} with the temperature given in reduced units. We see that although there are considerable variations in the shape of the curve, the maximum value of $\alpha_\parallel$ is rather similar in all three cases.

The longitudinal coefficient $\alpha_{\parallel}$ undergoes a sign change at a temperature of about 100 K, \cite{Astrov2} the origin of which remains unclear. Calculations of Ref.\ \onlinecite{Iniguez} gave a negligible value of the spin contribution to $\alpha_{\parallel}$ at zero temperature. It was further found \cite{Malashevich} that the orbital contribution dominates over the spin magnetism and has the right sign, but its magnitude is still too small compared to experiment.

As is clear from Eq.\ (\ref{gn}) and (\ref{alpha-el}), the spin contribution does not change sign if the magnetic interaction is of a purely Heisenberg type. However, non-Heisenberg contributions could, in principle, make the \emph{effective} parameter $\Delta$ depend on the order parameter and thereby on temperature. A typical term in the Hamiltonian capable of inducing such an effect is $K_{12,13}(\mathbf{S}_1\mathbf{S}_2)(\mathbf{S}_1\mathbf{S}_3)$, where 2 is a nearest and 3 a next-nearest neighbor of site 1. Although very likely small compared to $J_1$ and $J_2$ exchange parameters, $K_{12,13}$ could be comparable to the relatively small $J_4$ which largely determines $\Delta$. A sign change of the effective $\Delta$ would manifest itself as a sign change of the magnetoelectric coefficient. However, $K_{12,13}$ and other such terms do not contribute to $\Delta$ calculated, as we did, using collinear spin configurations.

\begin{figure}[htb]
\includegraphics*[width=0.45\textwidth]{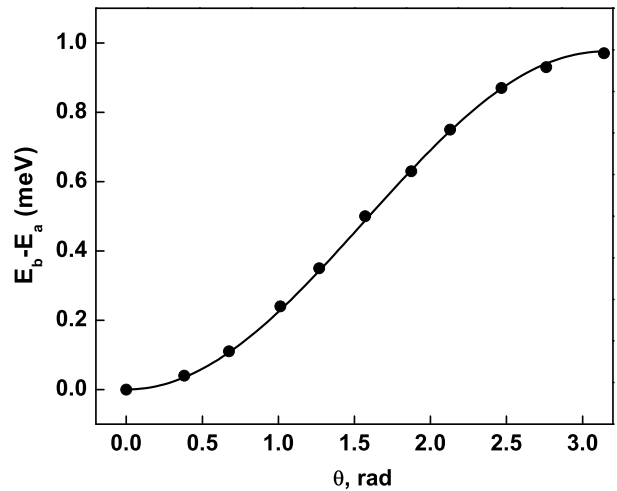}
\caption{Angular dependence of the parameter $\Delta=E_A-E_B$. The solid line shows the cosine fit to the data points (discs).}
\label{angular}
\end{figure}

To test whether non-Heisenberg terms like $K_{12,13}$ are appreciable in \CHR, we calculated the total energies $E_A(\theta)$ and $E_B(\theta)$ of a supercell with one Cr spin continuously rotated by an angle $\theta$ from 0 to $\pi$ on either of the two sublattices with ionic displacements induced by electric field. These calculations were performed using the self-consistently determined constraining fields (as implemented in VASP) but are otherwise similar to the evaluation of $\Delta$ (at $\theta=0$ and $\theta=\pi$ they are equivalent). The difference $\Delta(\theta)=E_A(\theta)-E_B(\theta)$ is plotted in Fig.\ \ref{angular}. It is seen that $\Delta(\theta)$ fits very well to a simple cosine. This indicates that the effect of non-Heisenberg interaction on $\Delta$ is negligible, and that the origin of the sign change in $\alpha_\parallel$ should be sought elsewhere.

\begin{figure}[htb]
\includegraphics*[width=0.45\textwidth]{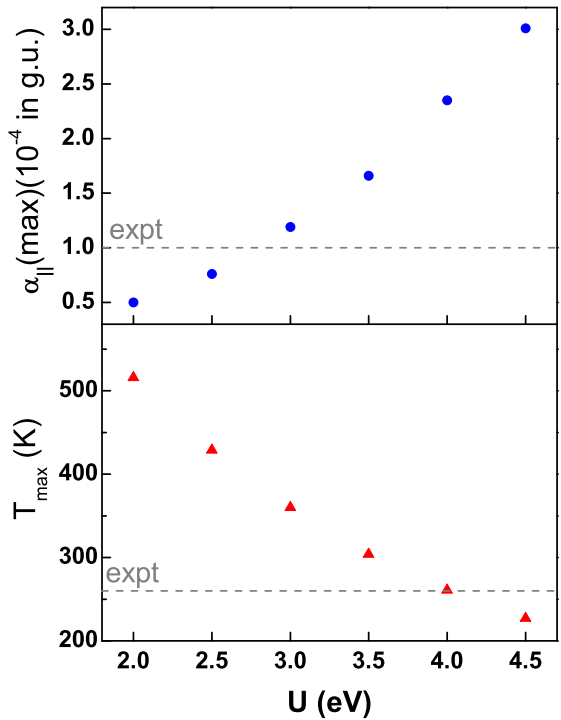}
\caption{$\alpha_{max}$ and $T_{max}$ calculated within the pair-cluster approximation as a function of the Hubbard $U$ parameter. The grey dashed lines show the corresponding experimental values.}
\label{Udep}
\end{figure}

\begin{figure}[htb]
\includegraphics[width=0.45\textwidth]{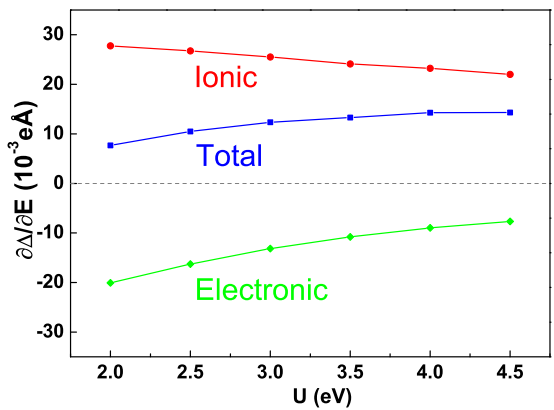}
\caption{Parameters $\partial\Delta/\partial E$ for lattice-mediated and electronic contributions as a function of the Hubbard $U$ parameter.}
\label{delta-U}
\end{figure}

The maximum value of $\alpha_\parallel$ ($\alpha_{max}$) obtained in the pair-cluster approximation and the temperature $T_{max}$ at which this maximum is reached are shown in Fig.\ \ref{Udep} as a function of the Hubbard $U$ parameter used in the calculation. We see that $\alpha$ increases by a factor of 2 when $U$ is increased from 3 to 4 eV. In order to understand the origin of this strong dependence, we first examine the $\partial\Delta/\partial E$ factors for the lattice-mediated and electronic contributions, which are shown in Fig.\ \ref{delta-U}. We see that $\partial\Delta/\partial E$ decreases as a function of $U$ for both lattice-mediated and electronic mechanisms. However, the reduction of the electronic term is faster, so the total value increases, albeit rather slowly. Thus, the overall strong dependence of $\alpha_{max}$ on $U$ is almost entirely due to the enhancement of the magnetic susceptibility.

While the peak temperature $T_{max}$ calculated in the pair cluster approximation agrees with experiment for $U=4$ eV, which also provides optimal description of the electronic and structural properties, \cite{Shi} the peak magnetoelectric susceptibility $\alpha_{max}=2.35\times10^{-4}$ g.\ u.\ is strongly overestimated compared to the experimetal data, which appear to converge to the value of about $1.0\times10^{-4}$ g.\ u. \cite{Astrov,Folen,Astrov2,Kita,Kita2,Rivera,Wiegelmann,Wiegelmann2,Borisov} It is also larger than the value obtained by Mostovoy \emph{et al.} who used a smaller value of $U$, treated the spins classically, and used a different method of extracting the magnetoelectric coupling constant. (If we use the same parameters and approximations, the results of Ref.\ \onlinecite{Mostovoy} are reproduced.) The main reason for the disagreement in $\alpha_{max}$ with experiment is in the overestimation of the magnetic susceptibility. The experimental value near the N\'eel temperature is $\chi(T_N)\approx 25$ emu/g, \cite{Foner} whereas the pair-cluster approximation yields 42 emu/g. If we use the experimental susceptibility instead of the calculated one, $\alpha_{max}$ comes out at about $1.4 \times10^{-4}$ g.\ u., which is much closer to the experimental values.

The reason for the overestimation of $\chi$ can be qualitatively understood using the Curie-Weiss expression $\chi=C/(T+\Theta)$, where $C$ is the Curie constant and $\Theta$ the Curie-Weiss temperature. For $S=3/2$, using the experimental value 25 emu/g for $\chi(T_N)$, we obtain $\Theta\approx450$ K. Similar values of $\Theta$ were found from the high-temperature susceptibility measurements \cite{Foex} ($550\pm50$ K) and from the exchange parameters obtained by fitting the inelastic neutron scattering data for magnon dispersions \cite{Samuelsen} ($527\pm76$ K). We can write $T_N\sim a \xi J_s$ and $\Theta\sim a |J_0|$, where $a$ is a common coefficient, $J_s=\sum_j\mathbf{e}_i\mathbf{e}_jJ_{ij}$, $J_0=\sum_j J_{ij}$, and $\xi$ is a suppression factor showing how much $T_N$ is suppressed by fluctuations compared to its mean-field value. In the pair-cluster approximation $\xi\approx0.8$. Thus, based on the experimental data we can conclude that $|J_0|$ is slightly greater than $J_s$. This implies that the intrasublattice exchange interaction is small compared to the intersublattice one. The results of our calculations contradict this picture, giving $|J_0|/J_s\approx0.42$ due to a fairly large value of $J_4\approx-0.2J_1$ at $U=4$ eV.

We have verified the fidelity of our fit of the exchange parameters to the calculated total energies by increasing the number of input configurations to 42. The results are listed in Table \ref{Jij}, which shows that the fit is quite stable. Particularly, the take-one-out cross-validation (CV) score for this five-parameter fit is 0.7 meV. If $J_4$ is not included in the fit, a much larger CV score is obtained. If an additional parameter $J_6$ is included, its value comes out an order of magnitude smaller than $J_4$. These results suggest that $J_4$ is too large due to the inaccuracies of the electronic structure in the LDA$+U$ method. It is known that LDA systematically underestimates the binding energy of the oxygen $2p$ states in oxide insulators. In \CHR\ this leads to an overestimated hybridization with the Cr $3d$ states, which tends to increase with increasing $U$ due to the downward shift of the filled $3d$ orbitals. Therefore, since $J_4$ is expected to be dominated by superexchange, its overestimation in LDA$+U$ is quite natural.

\begin{table}[htb]
\caption{Exchange parameters $J_n$ (meV) obtained by fitting to total energies calculated at $U=4$ eV. A long dash indicates that the corresponding $J_n$ is not included in the fitting. The cross-validation score (CV, meV) for each fit is also provided.}
\begin{tabular}{|c|c|c|c|c|c|c|}
\hline
$J_1$      & $J_2$    & $J_3$          & $J_4$      & $J_5$      &$J_6$    & CV \\
\hline
14.64      &  11.12    &    -2.11       & -2.98         &   2.12      &      ---    &     0.7        \\
19.57      &  14.18    &    -1.94       & ---              &   2.24      &      ---    &     6.9        \\
14.64      &  11.12    &    -2.12       & -2.98         &   2.12      &   0.12    &     0.1         \\
\hline
\end{tabular}
\label{Jij}
\end{table}

\section{Conclusions}\label{conclusions}

We formulated a microscopic model of the temperature-dependent exchange-driven magnetoelectric susceptibility $\hat\alpha$ which includes the coupling of scalar spin products to atomic displacements and to the electric field. The parameters of the model can generally be obtained using first-principles calculations, and it can be extended to magnetoelectric alloys, which may help in the search for new materials with better magnetoelectric properties. If the intrasublattice spin correlations can be neglected (which is a good approximation for Cr$_2$O$_3$), then $\alpha_\parallel$ can be expressed as a product of the magnetic susceptibility, sublattice magnetization, and a factor that does not depend on temperature, as long as the elastic properties of the lattice do not depend on it. This relation was suggested phenomenologically by Rado. \cite{Rado}  If, further, the intrasublattice interactions beyond the fourth coordination sphere in \CHR\ are negligibly small, our approach essentially becomes the converse of that of Mostovoy \emph{et al.}\cite{Mostovoy} Lattice-mediated and electronic contributions to $\alpha_\parallel$ have been sorted out, and the former was decomposed in the sum of contributions from the two normal displacement modes.

The electronic contribution to $\alpha_\parallel$ in \CHR\ is comparable to the lattice-mediated contribution and has an opposite sign. Quantum pair cluster and mean-field approximations for spin $3/2$, as well as the classical mean-field approximations result in similar peak values of the magnetoelectric susceptibility. The latter, however, is a quickly increasing function of the Hubbard $U$ parameter, mainly thanks to the increasing magnetic susceptibility $\chi$. If $\chi$ is taken from experiment, we find $\alpha_\parallel$ in good agreement with experiment. However, the calculation at $U=4$ eV, which results in a good agreement with experiment for many other properties, overestimates $\chi$ by a factor of 1.7, which in turn is due to the relatively large value of the $J_4$ parameter. Finally, it was found that non-Heisenberg exchange in \CHR\ is negligibly small and can not account for the sign change of $\alpha_\parallel$ observed at low temperatures.

\section{Acknowledgments }
This work was supported by the NSF/SRC Supplement to the Nebraska MRSEC (DMR-0820521), the Center for NanoFerroic Devices (CNFD) and the Nanoelectronics Research Initiative (NRI). Computations were performed utilizing the Holland Computing Center at the University of Nebraska.

\appendix

\section{Longitudinal displacement modes and phonon frequencies}

The normalized A$_{2u}$ eigenmodes of the force constant matrix $\hat A$ are listed in Table \ref{displace}. The frequencies of the infrared-active phonon modes in \CHR\ are compared with experimental data and with the results of Ref.\ \onlinecite{Iniguez} in Table \ref{frequency}.

\begin{table}[htb]
\caption{Normalized A$_{2u}$ eigenmodes of the $\hat A$ matrix.}
\begin{tabular}{|c|c|c|}
\hline
       &LO$_1$ & LO$_2$ \\
\hline
Cr$_1$  &$(0,0,0.361)$     &$(0,0,0.140)$      \\
Cr$_2$  &$(0,0,0.361)$     &$(0,0,0.140)$      \\
Cr$_3$  &$(0,0,0.361)$     &$(0,0,0.140)$      \\
Cr$_4$  &$(0,0,0.361)$     &$(0,0,0.140)$      \\
O$_1$   &$(0.074,-0.128,-0.241)$ &$(-0.190,0.330,-0.094)$    \\
O$_2$   &$(-0.148,0,-0.241)$     &$(0.381,0,-0.094)$    \\
O$_3$   &$(0.074,0.128,-0.241)$  &$(-0.190,-0.330,-0.094)$     \\
O$_4$   &$(0.074,-0.128,-0.241)$ &$(-0.190,0.330,-0.094)$    \\
O$_5$   &$(-0.148,0,-0.241)$     &$(0.381,0,-0.094)$    \\
O$_6$   &$(0.074,0.128,-0.241)$  &$(-0.190,-0.330,-0.094)$    \\
\hline
\end{tabular}
\label{displace}
\end{table}

\begin{table}[htb]
\caption{Table lists all calculated polar phonon frequencies (in unit of cm$^{-1}$), and compare with experimental data\cite{Lucovsky} and \'I\~niguez's DFT calculation. \cite{Iniguez}}
\centering
\begin{tabular}{|c|cc|cccc|}
\hline
                                                    & \multicolumn{2}{c|}{$A_{2u}$ modes }     & \multicolumn{4}{c|}{$E_u$ modes} \\
\hline
      This work                                     &   572   & 409                                            & 637   & 567 & 450 & 312   \\
      Experiment (Ref.\ \onlinecite{Lucovsky})      &   533   & 402                               & 609  & 538 & 440 & 305    \\
      Other theory (Ref.\ \onlinecite{Iniguez})     &   597   & 408                                            & 653  & 578 & 455 & 316    \\
\hline
\end{tabular}
\label{frequency}
\end{table}

\end{document}